\documentclass[aps,10pt,prb,twocolumn,showpacs,showkeys,citeautoscript]{revtex4-1}
\usepackage[utf8]{inputenc}

\usepackage{graphicx}
\usepackage{amssymb}
\usepackage{amsmath}

\begin{document}
\title{Role of nonlinearity in resistive switching phenomena of lanthanum calcium manganate heterostructures}
\author{I.V. Boylo}
\affiliation{Donetsk Institute for Physics and Technology, Rosa Luxembourg Str.~72, Donetsk, Ukraine 83114}
\email{boylo@fti.dn.ua}
\date{\today}
\begin{abstract}
The present paper examines the influence of a nonlinear relationship between
the local oxygen-vacancy concentration and the local resistivity 
on resistive switching effects in complex oxides.
The continuity equation has been used as a model for the
motion of oxygen vacancies when a periodic time-dependent electrical current
is applied. The question of endurance of the switching cycles is discussed.
It is found that nonlinearity of the resistivity-concentration dependence
enhances the endurance.
\end{abstract}
\maketitle

\section{Introduction}

It has been recently shown \cite{Tang2016} that the shock-wave formation strongly
influences resistive switching effect in manganese oxides.
These authors first pointed out that it is a shock wave of 
oxygen vacancies that  provides
the change of the resistance.

Resistive switching was discovered in 1960s in oxides sandwiched between metal 
electrodes \cite{hickmott1962,nielsen1964,gibbons1964}. 
It had been realized that this effect was promising for 
non-volatile memory applications \cite{nielsen1964}.
Since then resistive switching phenomena
has puzzled a lot of scientists. 
The burst of scientific activities begun in 2008 when two-terminal devices that demonstrate
hysteretic current–voltage behavior were understood as memristor devices \cite{strukov2008}.
 Nowadays, the resistive switching effect is one of the best-known phenomena 
in complex oxide-based heterostructures proposed for a novel memory cell.
A lot of scientists and engineers are engaged in studies
and applications of the resistive switching. 
Presently, some plausible models for resistive switching in
oxides have been reported\cite{Waser2007,Sawa2008,lee2015}. 
Among them is the voltage-enhanced 
oxygen-vacancy migration model (VEOVM) for
bipolar resistive switching\cite{Rozenberg2010}.

An essential role of oxygen vacancies and 
their motion on the resistive switching effect in transition-metal oxides 
has been established in last decades \cite{Waser2007,Sawa2008,Bryant2011}.
In particular, oxygen diffusion is considered to be a key factor of the switching in
Pr$_{0.7}$Ca$_{0.3}$MnO$_3$ with Schottky-like barrier at the interface
\cite{Nian2007,Sawa2008}.
The experimentally observed asymmetry of
current-voltage characteristics 
has been recently explained using the oxygen diffusion scenario in yttrium barium
copper oxide \cite{PT2012}. Rearrangements of oxygen vacancies diffusing to and away
from metal-complex oxide interfaces
 can although correlate with the evolution of the resistance of the La$_{0.5}$Ca$_{0.5}$MnO$_3$ based
 memristive device \cite{Wang2012}.

In this paper, we chose lanthanum calcium manganate to model
the resistive switching phenomenon in metal-complex-oxide-metal heterostructures.
The doped manganese oxide La$_{0.7}$Ca$_{0.3}$MnO$_3$ (LCMO) 
is attractive due to its benefits for practical applications at room temperature, it is 
already being actively investigated \cite{shang2006,dong2007}. 
It was shown that the LCMO-based memristive device can demonstrate
good endurance of switching cycles \cite{yang2010}.

It is known that the oxygen nonstoichiometry of complex oxides
affects their resistivity. For instance, a shift of the metal-to-insulator transition
as a function of the oxygen deficiency ($\delta$) in perovskite-type
La$_{0.7}$Ca$_{0.3}$MnO$_{3-\delta}$ 
thin films exists in the range of 0$< \delta<$0.21\cite{Rubio-Zuazo2014}.
 The film-substrate interface can influence the resistive properties of 
epitaxial La$_{0.7}$Ca$_{0.3}$MnO$_3$ thin films as well \cite{Lu2000}.

Previous studies assumed that there is a linear relationship
 between the local resistivity and the oxygen-vacancy concentration
 in manganites \cite{Rozenberg2010,rozen2010,Tang2016}. 
  However,  in reality the form of such dependence is more complicated.
  For example, the local electrical resistivity of
  transition-metal oxides with a perovskite-type structure 
  as a nonlinear function
  of the oxygen-vacancy concentration was obtained experimentally
  in a wide temperature range \cite{Malavasi2004,Baskar2008}.
 As was pointed out, the type of such dependence can strongly affect
 the hysteresis observed
 in metal-oxide heterostructures \cite{Zhao_Liang2013}.
 The goal of this paper is to explore how it controls the resistive switching in
 La$_{0.7}$Ca$_{0.3}$MnO$_{3-\delta}$ .

 \section{Model}
 
 We follow the idea that the formation of a shock wave
 provides a sharp change in the resistance of manganese oxides \cite{Tang2016} and 
 start with the one-dimensional continuity equation for mobile oxygen vacancies
\begin{equation}
\label{eq:continuity eq}
\partial_t c\left(t,x\right) + \nabla j\left(t,x\right)=0,
\end{equation}
where the local concentration of mobile vacancies $c\left(t, x\right)$ changes in time $t$,
and the total oxygen-vacancy current
$j\left(t,x\right)$ is divided in two parts:
 a diffusion current $j_\mathrm{diff}$ and a drift
current $j_\mathrm{drift}$, so that $j\left(t,x\right)=j_\mathrm{diff}+j_\mathrm{drift}$.

The flow of vacancies from high to low concentrations along the space coordinate $x$ 
normal to the cross-section of the LCMO-based heterojunction
can be described
by the Fick's first law of diffusion $j_\mathrm{diff}=-D \partial_x c$, while
 the electric field $E$ is involved in the resistance switching 
 as the driving force for the drift 
 $j_\mathrm{drift}=c\mu E$. 
 We note that according to the Einstein relation 
  the mobility $\mu$ of the moving charge $q$  is connected with the diffusion constant $D$ via $\mu =qD/ (k_\mathrm{B} T)$,
  where $k_\mathrm{B}$ is the Boltzmann constant, and $T$ is the absolute temperature in kelvins. 
 
 The probability for oxygen vacancy to overcome the energy barrier and
 jump back ($\leftarrow$) or forth ($\rightarrow$) 
by the distance $a$ 
between any two neighbouring lattice sites is 
\begin{displaymath}
r_{\rightleftarrows}=\frac{1}{2} \nu \exp \left(-\frac{E_A \mp \Delta E}{k_\mathrm{B} T}\right)
\end{displaymath}
with $\nu $ being the attempt frequency, and
$E_A$, the activation energy for vacancy motion.
Due to the activation-barrier lowering $\Delta E=qaE$ caused by the electric field
$E$ applied to the metal electrodes oxygen vacancies in LCMO are mobile enough 
to migrate detectably at room temperature\cite{warnick2011}. 
 
The average velocity $v= \mu E$ of such motion is given by
\begin{displaymath}
v=a \left(r_{\rightarrow}-r_{\leftarrow}\right)=a \nu
\exp \left(-\frac{E_A}{k_\mathrm{B} T}\right)
\sinh \left(\frac{qaE}{k_\mathrm{B} T}\right).
\end{displaymath}

 We restrict ourselves to the case of small electric fields $qaE\ll k_\mathrm{B} T$ when
the average mobility of the oxygen vacancies is
\begin{equation}
\label{eq:mobility}
\mu=\frac{qa^2 \nu}{k_\mathrm{B} T} 
\exp \left(-\frac{E_A}{k_\mathrm{B} T}\right).
\end{equation}

The connection between the local electric field and the electrical current $I$ is determined by the Ohm’s law 
 $E=\rho \left(\delta \right) I$, where 
 the resistivity $ \rho $ is some function of 
 the oxygen-vacancy concentration. 
In La$_{0.7}$Ca$_{0.3}$MnO$_{3-\delta}$ compounds, 
 as can be extracted from the 
 electrical resistivity versus temperature measurements
 for the samples with the oxygen deficiency $ \delta $ = 0, 0.01, 0.025,
 the relationship between the local resistivity $ \rho$ and 
 the oxygen-vacancy concentration $\delta$
 is a nonlinear one \cite{Malavasi2004}:
\begin{equation}
\label{eq:localresistivity}
\rho \left(\delta\right) = \rho_0 ( 1+\alpha_{1}\delta - \alpha_{2} \delta^2 ),
\end{equation}
where the $\rho_0 $, $\alpha_{1}$, and $\alpha_{2}$ are constants derived from 
the experimental data.

Note that the effect of oxygen vacancies
in metallic LCMO is negligible and the resistivity tends to zero $\rho_\mathrm{metal}=0$.
So we can write the total content of oxygen vacancies $\delta$ as $c+c_\mathrm{metal}$, 
where $c$ has the meaning of 
the concentration of mobile oxygen vacancies, and  
$c_\mathrm{metal}$ is the vacancy concentration at which the metal-insulator transition occurs.

The diffusivity $D$ is usually of the form $D=D_0 \exp \left(-E_A/k_\mathrm{B} T\right)$, where
$D_0=a^2 \nu /2$ being the frequency factor.
Therefore, the drift of diffusing species with the average mobility (\ref{eq:mobility}) 
is given by  
$$j_\mathrm{drift}=\frac{2qD}{k_\mathrm{B} T} c \rho \left(c \right) I.$$
To include the possibilities for non-harmonic driving current $I$, by which oxygen vacancies
are transported from one site to another, let us model its time dependence as
$I(t)=I_\mathrm{max} \mathrm{sn}(4 (t/T_0) \mathrm{K}(m) | m)$, where $\mathrm{sn}(u|m)$
is Jacobi elliptic 
sine function and $\mathrm{K}(m)$ is the complete elliptic integral of the first kind. 
This function is periodic with period $T_0$ and its shape can be continuously varied 
between the sine-wave (at $m=0$) and the square-wave (at $m=1$).

Based on the continuity equation (\ref{eq:continuity eq}),
migration of the charged species in
a thin film of manganese oxide with a thickness $d$
normal to the metal/LCMO interface can be modeled by the equation
\begin{equation}
\label{eq:nonlinear diffusion}
\partial_t c+\frac{2qDI(t)}{k_\mathrm{B} T} \partial_x \left(c\rho (c ) \right)= D\partial_{xx} c,
\end{equation}
in which the first term tells us that the oxygen-vacancy content changes in time,
the second term represents a drift current of charged species, 
and the third one describes diffusion driven by the concentration gradient.

On introducing new scales for variables
$$
t \rightarrow t /T_0, 
x \rightarrow x /d,
$$
and $I(t) \rightarrow I(t)/I_{max} $ such that $I_{max}$ being the maximum value
of the electrical current applied,
the nonlinear equation  (\ref{eq:nonlinear diffusion})  becomes a dimensionless one:
\begin{equation}
\label{eq:dimensionlesseq}
\partial_t c+2 \gamma_1 \gamma_2 I\left(t\right) 
f\left(c\right) \partial_x c= \gamma_1 \partial_{xx} c,
\end{equation}
where 
$$
f \left(c \right)=1+ \alpha_{1} \left(2c- c_\mathrm{metal} \right)-
\alpha_{2} \left((2c-c_\mathrm{metal})^{2}-c^{2} \right),
$$
$\gamma_1 = DT_0/d^2$, 
$\gamma_2 =dq\rho_0 I_{max}/k_\mathrm{B} T$, 
and the diffusion coefficient $D$ can assumed constant at the given temperature.

Note that if $f\left(c\right)\simeq c$, then the equation (\ref{eq:dimensionlesseq}) reduces to
the well known nonlinear 
Burgers’ equation. For small enough diffusion it has shock wave 
solutions in the form of an abrupt step (that is a jump) in concentration.

Consider the uniform distribution of the oxygen vacancies
$c_0$ initially. 
As it was noted \cite{dong2007,Liu2010}, the system of interest is able to switch continuously  
 between two resistive states as long as there is a source of oxygen vacancies
 at the metal/LCMO interface. 
 Therefore, we suggest that the concentration 
 at the $x=d$ surface of the LCMO is maintained at the initial value,
 $c_0$, due to surface oxygen-exchange processes
\cite{Bryant2011}, while surface flux across the $x=0$ interface 
between  the  substrate  and  the  LCMO is assumed to be zero
(cause, in practice, there is no flow of the moving species through it).
This can be expressed as boundary conditions
\begin{equation}
\label{eq:boundary conditions}
\left. j\left(t,x\right)\right|_{x=0}=0,
\mathrm{but} \left. c\left(t,x\right)\right|_{x=d}=c_0.
\end{equation}

When the electric current is supplied, oxygen vacancies are driven backwards or forwards
depending on its polarity.
Such a motion is strongly coupled
with the resistivity of the complex oxide.
To elucidate the effects of this coupling on the transport properties, the simplest approach
is to solve the nonlinear diffusion equation
(\ref{eq:dimensionlesseq}) numerically.

\section{Results and discussion}

Let us investigate the Eq. (\ref{eq:dimensionlesseq}) 
with the boundary conditions (\ref{eq:boundary conditions}) to establish the interdependence between 
the concentration profile in LCMO thin film and the switching of the contact resistance. 
This requires the values of dimensionless parameters $\gamma_1$ and $\gamma_2$ to be specified.
The parameter $\gamma_1$ incorporates the thin-film thickness $d$, the 
diffusion coefficient $D$ of the medium, as well as the period of the 
electrical signal $T_0$. To make the specific calculations we assume
$\gamma_1= 0.015$, which gives us stable switching behavior. The parameter
$\gamma_2$ is estimated from the condition that the resistive switching
effect at room temperature in the lanthanum calcium manganate
is typically observed\cite{dong2007,yang2010,Liu2010} at the voltage of about $2 \mathrm{V}$
between the electrodes,
which yields $\gamma_2=160$. The initial concentration of mobile oxygen 
vacancies in the manganese oxide (and on the boundary $x=d$) is fixed to be
constant, $c_0=0.004$ (not very large, which ensures that we do not exceed the
range of experimental $\rho(\delta)$ values).

\begin{figure}[htbp]
 \includegraphics[width=\columnwidth]{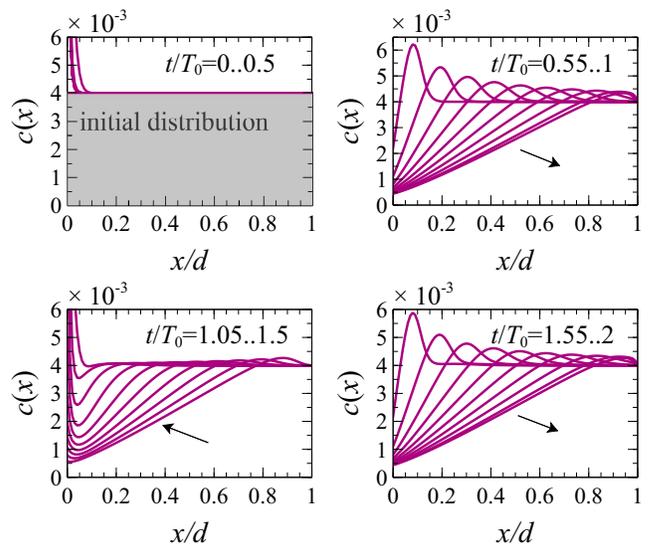}
 \caption{\label{fig:concentration}Concentration profiles of oxygen vacancies at different times. 
 Panels correspond to four half-periods: $t/T_0=0..0.5$,
$t/T_0=0.55..1$, $t/T_0=1.05..1.5$,
$t/T_0=1.55..2$; arrows indicate the direction of the wave movement.}
\end{figure}

Spatial distribution of the oxygen-vacancy concentration
is sampled with equal time step of $0.05 T_0$ in Fig.~\ref{fig:concentration}.
Initially at $t=0$  (top-left panel) the concentration
profile is a constant, the even half-periods (right panels) show the development of a shock-wave profile
forming at times $t/T_0=0.5;1.5$. Notice that at times $t/T_0=0..0.45;1.05..1.45$ (left panels), 
vacancies pile up at the metal/LCMO interface.
The vertical range in the left graphs is reduced to show the formation of the shock-wave front at times 0.5;1.5, 
otherwise the concentration at $x=0$ can be as high as $46*10^{-3}$.

The interface $x=0$ (where most changes occur) is nontransparent for
oxygen vacancies with zero-flux boundary conditions (\ref{eq:boundary conditions}).
So, the positively charged oxygen vacancies drift to that interface
and accumulate
at the metal/LCMO interface (left panels in Fig.~\ref{fig:concentration}).
After changing the polarity of the electrical signal, 
the oxygen vacancies travel back from the surface into the bulk of the manganese oxide,
and the solution
of the equation  (\ref{eq:dimensionlesseq})
 transforms into some form of a steep shock wave 
 (right panels in Fig.~\ref{fig:concentration}),
 and the final traveling wave profile at $t/T_0=1;2$. 
  
 Now let us discuss the relationship 
between the oxygen-vacancy distribution and the corresponding changes in the resistance.

The total resistance $R$ for a film between $x=0$ and $x=d$
 can be computed as
  \begin{equation}
  \label{eq:contact resistance}
R(t)=\int_{0}^{d} \rho \left( c (t, x) \right) {\mathrm{d}}x,
\end{equation}
where the local resistivity $ \rho$,
which is a function of the local vacancy concentration $\delta$ or local mobile 
vacancy concentration $c$,
is given by the Eq. (\ref{eq:localresistivity}).
According to the experimental data \cite{Malavasi2004} the 
constant $\rho_0$ for the sample with $\delta=0$ is
$\rho_0=0.37$ $\Omega \rm{cm}$. 
\begin{figure}[htbp]
 \includegraphics[width=\columnwidth]{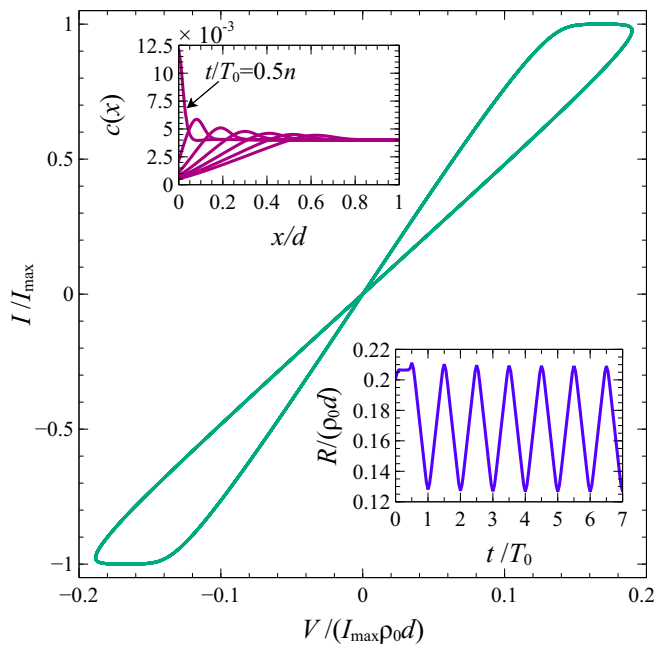}
 \caption{\label{fig:Resistance_concentration} 
 Electrical current-versus-voltage ($I$-$V$)
 characteristic
in units of $I_{\mathrm{max}}$ and $I_{\mathrm{max}} \rho_{0}d$, respectively. 
Insets: propagation of the oxygen-vacancy wave (top inset), 
and the corresponding resistance $R(t)$ curve (bottom inset); $n$ is an integer.}
\end{figure}

The current-voltage ($I$-$V$) characteristics, shown in Fig.~\ref{fig:Resistance_concentration},
 represent the sharp switching between the high-resistance state and the low-resistance
state. The top inset of Fig.~\ref{fig:Resistance_concentration} shows the shock-wave formation
of the oxygen-vacancy concentration at $t/T_0=0.5n$.
 The bottom inset depicts the corresponding time evolution of the resistance
 calculated using
 Eq.(\ref{eq:contact resistance}).
 After the first forming cycle 
the total resistance of the heterostructure changes periodically.
In our numerical calculations, these reversible changes between the two resistive states occur
up to fifty times
(which is not shown in Fig.~\ref{fig:Resistance_concentration}) that proves the stability
of switching.
Initially, the resistivity of the junction is high due to the presence
of some amount of oxygen vacancies in the equilibrium doped manganese oxide. It is then somewhat 
increased in the forming process. 
As the oxygen vacancies leave the interface, the resistance decreases.
We can also see that the rise in
oxygen deficiency at the $x=0$ interface 
brings the system into a high resistive state. 

Thus, the oxygen-vacancy species
 moving forwards and backwards can produce the sharp increase or decrease
 in the resistance, but also the way the oxygen-vacancy concentration profile is changing
 tells us about the formation of the shock wave within
 the nanoscale layer at the nontransparent metal/LCMO interface.
The moment $t/T_0=0.5n$, when the concentration wave changes its profile,
directly corresponds to the sudden switching 
of the resistance of the sample, and
it can be seen clearly
how each time the resistivity and the oxygen-vacancy distribution change together. 

 Obviously, a sudden switching of the resistance is an unavoidable feature of 
Eq. (\ref{eq:dimensionlesseq}),
mediated by the shock-wave profile of the vacancy concentration.
But the relevant feature is that, as shown in Fig.~\ref{fig:Resistance_concentration}, 
the system demonstrates reproducible switching between the
two resistive states after only a few electrical cycles. 
It is the effect of the nonlinear term entirely.

\begin{figure}[htbp]
 \includegraphics[width=\columnwidth]{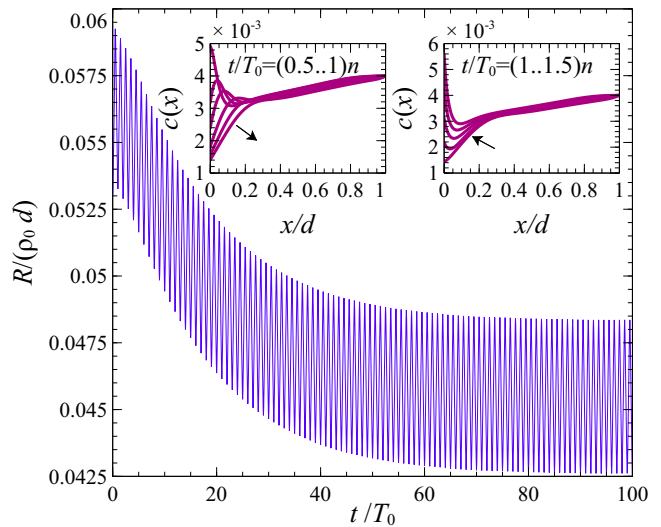}
 \caption{\label{fig:Resistance_concentration_Line}Time dependence of the contact resistance $R$ in 
 units of $\rho_{0}d$ versus dimensionless time $t/T_{0}$. Inset: time evolution of the depth profile
 of the oxygen-vacancy concentration at times $t/T_0=(0.5..1)n$ and $t/T_0=(1..1.5)n$, respectively.}
\end{figure}

To obtain a linear function $\rho \left( \delta \right)$ from the data\cite{Malavasi2004}
we fit the dependence of the local resistivity on the oxygen-vacancy content\cite{Malavasi2004} 
by 
 \begin{equation}
\label{eq:localresistivity_Line}
\rho \left(\delta\right) = \rho_0 ( 1+\beta\delta ).
\end{equation}
Using this function in Eq. (\ref{eq:nonlinear diffusion}) results in the concentration profiles,
shown in the inset to Fig.~\ref{fig:Resistance_concentration_Line}.
As can be seen, abrupt jumps in the vacancy content
also occur at the interface. 
Meanwhile, in the bulk of LCMO, the concentration remains nearly
constant, which
implies a much smaller switching of the total resistance.

We can see in Fig.~\ref{fig:Resistance_concentration_Line}
that the sudden changes between
 the high-resistance state and the low-resistance
state is observed.
However, there are significant changes in
both the local concentration and the contact resistance behavior in the sample.
It looks like an excellent switching -- very reversible and stable, but the stability
of the contact resistance in high and low resistance states is only achieved after about 80 time periods.

\begin{figure}[htbp]
 \includegraphics[width=\columnwidth]{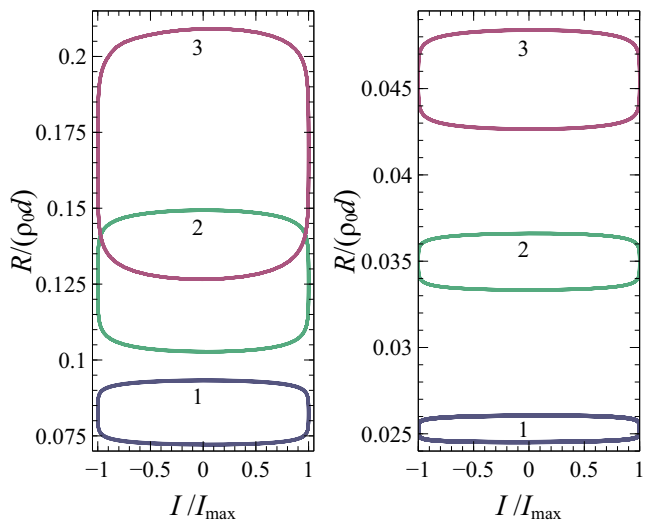}
 \caption{\label{fig:Resistance vs current} Comparison of resistance-versus-electrical current ($R$-$I$)
hysteresis loops in units of $\rho_{0}d$ and $I_{\mathrm{max}}$
between the cases of nonlinear dependence $ \rho \left( \delta \right)$
after skipping a few cycles of the electrical current signal (left) and the case of linear dependence 
$ \rho \left( \delta \right)$
after skipping 80 cycles of the electrical signal (right). The initial mobile-vacancy content
$c_0$=0.002 (1), 0.003 (2), 0.004 (3), respectively.}
\end{figure}

\begin{figure}[htbp]
 \includegraphics[width=\columnwidth]{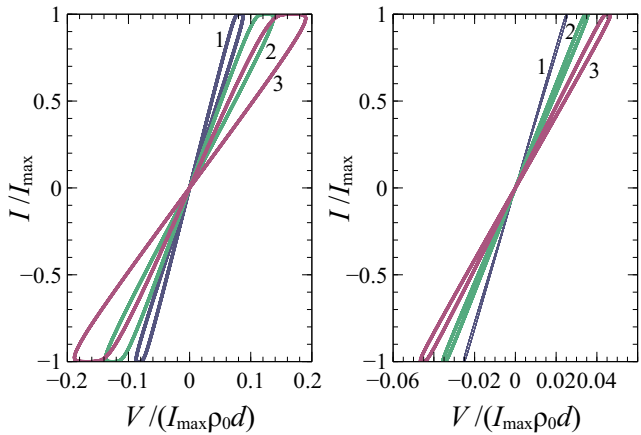}
 \caption{\label{fig: Current vs voltage}Electrical current-versus-voltage ($I$-$V$)
 hysteresis loops
in units of $I_{\mathrm{max}}$ and $I_{\mathrm{max}} \rho_{0}d$. 
The simulation parameters are the same as in Fig.~\ref{fig:Resistance vs current}.}
\end{figure}

In Figs.~\ref{fig:Resistance vs current} and ~\ref{fig: Current vs voltage}
we plot the hysteresis loops corresponding to three
different values of the initial concentration $c_0=0.002,0.003,0.004$.
These hysteresis loops are sensitive to the initial conditions both
in the case of linear (right)
and nonlinear (left)
dependence $ \rho \left( \delta \right)$: the larger is the mobile-vacancy content, 
the more significant is the switching effect.

The electrical resistance-versus-electrical current ($R-I$) 
hysteresis loops for LCMO-based heterostructures are shown
in Fig.~\ref{fig:Resistance vs current}. 
Even if the initial concentration of mobile oxygen vacancies $c_0$ is small,
reproducible $R-I$  hysteresis windows are clearly visible.
Each curve in Fig.~\ref{fig:Resistance vs current}
actually consists of five loops, 
which illustrates
high reversibility of the switching.
When $\rho \left( \delta \right)$ is linear, the $R$-$I$ characteristics show a little window.
In the case of nonlinear $\rho \left( \delta \right)$,
the shape of $R \left(I \right)$ dependence does not change, 
but the difference between high and low resistance is significantly larger.

In the two cases described above the current-voltage ($I$-$V$) characteristics also display hysteresis
 (see Fig.~\ref{fig: Current vs voltage}).
Even if the relationship between the local resistivity and the vacancy content
is linear some hysteresis occurs.
If nonlinear term exists in the function $\rho \left( \delta \right)$, 
the stable hysteresis takes place.

\section{Conclusions}

The first question for electronic applications of the resistive switching phenomena in
complex oxide-based devices
is how to stabilize the switching process by getting a good endurance of switching cycles.
To make switching behavior reliable, experiments have to be designed by fine tuning
the set of constants - the film thickness $d$, the 
diffusion coefficient $D$ of the medium, the period of the 
electrical signal $T_0$ as well, which should correspond
 to specific parameter $\gamma_1$.
If the relationship between
the local oxygen-vacancy concentration and the local resistivity is linear, 
we have to wait a long time for switchings become reversible
(in the case of interest - about 80 cycles of the electrical signal). 
The main finding here is that effect of nonlinear dependence $\rho (\delta)$ on hysteresis 
is such that
switchings become reversible after few cycles of the electrical signal.
Furthermore, nonlinear behavior of the local resistivity leads to
the larger hysteresis windows than in the case of linear dependence.

 Another result is that the sudden switching of the contact resistance
 is accompanied by the development of the shock-wave profile that have a
 common time $t/T_0=0.5n$.
 The shock-wave formation at the metal/LCMO interface has been captured 
 in the numerical solutions.
  For this reason, we can conclude that the resistance switching effect in manganese oxides
is correlated simultaneously with the shock wave formation and propagation, 
produced by the flow of oxygen vacancies from the nontransparent interface into the bulk of LCMO.
 On the contrary, in the case of linear fit $ \rho \left( \delta \right)$, 
the traveling concentration wave does not propagate across the whole bulk and
the hysteresis loses its endurance. This suggests that the design of the nonlinear
dependence of the resistivity on the vacancy concentration is necessary to obtain
strong and reliable resistance switching.

%
\end{document}